\documentclass[runningheads]{llncs}
\usepackage{graphicx}
\usepackage{tabularx}

\usepackage{amsfonts}

\usepackage{fancyhdr}

\fancypagestyle{reststyle}
{
		
	\fancyhf{}
	\fancyfoot[L]{\footnotesize Preprint - under review}
}

\begin{document}
\title{Drug Similarity and Link Prediction Using Graph
Embeddings on Medical Knowledge Graphs \thanks{Preprint - under review.}}
\titlerunning{Drug Similarity and Link Prediction on Medical Knowledge Graphs}
%
\author{Prakhar Gurawa \and Matthias Nickles \\p.gurawa1@nuigalway.ie, matthias.nickles@nuigalway.ie}
\authorrunning{P. Gurawa and M. Nickles}
\institute{School of Computer Science, National University of Ireland Galway}

\maketitle              
\begin{abstract}
The paper utilizes the graph embeddings generated for entities of a large biomedical database to perform link prediction to capture various new relationships among diﬀerent entities. A novel node similarity measure is proposed that utilizes the graph embeddings and link prediction scores to find similarity scores among various drugs which can be used by the medical experts to recommend alternative drugs to avoid side effects from original one. Utilizing machine learning on knowledge graph for drug similarity and recommendation will be less costly and less time consuming with higher scalability as compared to traditional biomedical methods due to the dependency on costly medical equipment and experts of the latter ones. 

\keywords{Knowledge Graphs \and Graph Embeddings \and Link Prediction \and Drug Similarity}
\end{abstract}

\section{Introduction}
Knowledge graphs~\cite{kg} are knowledge bases which represent domain knowledge 
as interlinked entities, forming the nodes of a graph. Driven by the recent ‘explosion’ 
of data, many corporations and academic institutions are relying on knowledge graphs
for the modelling and analysis of large amounts of data.
In this work, we use graph embeddings [7,12] to predict possible relationships
between drugs in a drug knowledge base represented as a knowledge graph. Finally, we intend to rely on graph embeddings to provide relevant information and predictions on the given database, as well as to assist us in performing the tasks of relation predictions using link prediction and drug-drug similarity utilising the node similarity concepts mentioned in this work.
Because of the reliance on expensive medical equipment and medical professionals to deal with nuances of the equipment and other activities, the field of drug similarity and analysis is time consuming and expensive. We aim to achieve very good results by employing knowledge graphs to develop drug-similarity and predictions in less time and at a lower cost than previous techniques. Our drug similarity model developed will help in drug similarity discovery which will help reduce the side effect~\cite{side-effect} caused by the use of alternative similar drugs.

\section{Related Work}
Currently, knowledge graphs~\cite{kg} are able to support many medical applications, due to the fact that graphs are a practical resource of many real-world applications~\cite{bio-med}. Typically, biological knowledge graphs are built using manually selected datasets such as MIMIC-iii, ICD-9, and others. Other options include using natural language processing to lessen the work of manual information collection. Today, knowledge graphs~\cite{kg} are used in a variety of biomedical applications such as Genomics, Proteomics, Drug Side Effects, Drug Repurposing, Safe Drug Recommendation, and many more, indicating their popularity in this field.
The research~\cite{rajeev} explores using representation learning in knowledge graphs for the study of drug target predictions and drug-drug interactions and~\cite{nuig} uses knowledge graph embeddings to perform link predictions and drug target discovery that too using the KEGG database. The work done in research~\cite{LSTM} utilizes the use of LSTM and knowledge graph embeddings based model to predict drug-drug interaction. 
A section of our work inspired from the works like~\cite{Distmult} which deals with making inferences using the learned models. The research explore various graph embedding model which has been described and compared in~\cite{KGE},~\cite{kg-survey} and~\cite{kg-survey2}.

\section{KEGG Database}
KEGG (Kyoto Encyclopedia of Genes and Genomes)~\cite{kegg}  is a knowledge base that contains genetic, chemical, and functional information. It contains various entities such as disease, gene, network, route, drug, and so on, and each entity type is linked to the others via a certain form of relationship as shown in the image below. KEGG was developed by Kanehisa Laboratories\footnote{Kanehisa Lab Website https://www.kanehisa.jp/} and is structured as a network of interconnected entities that resembles the biological ecosystem at the molecular level~\cite{nuig}. The database compromises ﬁve types of entities. The first one is drugs which are comprehensive drug information site that exclusively includes pharmaceuticals that are approved in Europe, Japan, and the United States, includes information about drugs such as molecular interactions, drug metabolism, and chemical structure. The second is genes which involves an amalgamation of genes and proteins, contains information about gene sequences and their interaction with other biological entities~\cite{nuig}. The next are diseases, collection of single-gene disorders, multifactorial diseases, and infectious diseases with a focus on perturbation, which deals with disease interactions with other entities. Pathways comprises manually selected pathway maps containing data on metabolisms, biological processes, human diseases, drug development, and other topics. Each route is linked to entities such as diseases, medications, and genes~\cite{nuig}. By linking distinct entities in other KEGG databases based
on this network database, the KEGG Network database represents information about medications and disorders in the form of molecular networks.

\section{Methodology}
\subsection{Knowledge Graph and Graph Embeddings}
A knowledge graph is defined as 
\emph{”a graph of data intended to accumulate and convey knowledge of the real world, whose nodes represent entities and whose edges represent relations between these entities.”}~\cite{kg}. Formally, a knowledge graph consists of triplets, consisting of two
entities (head, tail) connected or related using edges (relationship). The triplets are represented as (head,relationship, tail) or (h,r,t) with
h,t $\in$ E, r $\in$ R where E is a set of entities and R represents set of all possible relationships that can exist between any two entities. For example, KEGG contains instances like (D11034, DRUG\_EFFICACY\_DISEASE, H00409) which represents the drug and disease are connected by the specific relationship in the database.
Embedding learning is an efficient way to tackle data sparseness by representing knowledge graph’s entities and relationship as low dimensional real value vectors storing the structural characteristics of the graph structure within themselves~\cite{combine}. In research~\cite{kg-survey}, the authors have divided the graph embedding models in three families, the first translation distance based which utilizes distance based scoring methods, the second ones are semantic matching based which rely on similarity based scoring methods and finally the third family of neural network based models. 

\subsection{Link Prediction}
The task of predicting the connection between any two nodes based on their node properties is referred to as link prediction. The link prediction task in graph G, where each node represents some information, is to develop a model that predicts whether two nodes are related by an edge or not~\cite{linkpred0,linkpred}.This can be used to find additional information for an existing dataset and can also be considered an information extraction task. A link prediction job on a drug-drug interaction knowledge network, for example, can assist us in relieving new information by suggesting probable predictions between any new pair of medications.
For the task of link prediction, where every entity is considered as a target entity for a triple in testing data~\cite{rajeev}, metrics such as Mean Reciprocal Rank (MRR) and Hits@n are considered for evaluations. These scoring methods are then used to generate sorted scores based on corrupted or correct triplets~\cite{kgsim}.

\begin{itemize}
  \item MRR: Mean Reciprocal Rank calculates the mean of the reciprocals of vectors of ranking ranks. Also, MRR is less sensitive to outliers as compared to Mean Rank~\cite{transe}. MRR scores are standardised from 0 to 1, with 1 representing perfect ranking. \begin{equation}MRR = \frac{1}{|Q|}\sum\limits_{i=1}^Q \frac{1}{rank_i}\end{equation}
  Here, ${rank_i}$ refers to the rank position of the first relevant element
  for the i-th query~\cite{nuig}.
  \item Hits@n: It represents the model's likelihood of ranking the relevant (true) fact in the top k element scores in the rank by reporting how many elements of vectors of ranking rank made it to the top n forecasts~\cite{nuig}.
\end{itemize}

\subsection{Graph Embedding Algorithms}
The key characteristic of graph embeddings is that they hold the complex graph structure and interactions within themselves, with the distance between latent dimensions representing a metric for similarities between distinct graph elements~\cite{deepwalk}. The following graph embedding models has been used:

\begin{itemize}
\item TransE: An additive model which uses distance based scoring function for the link prediction task, treating edges of graph as linear transformations~\cite{transe}. The scoring function calculates a similarity between embedding of the head translated by the embedding of relationship and the embedding of the tail. Mathematically the scoring function is defined as below which involves using the $L_1$ or $L_2$ norms:
\begin{equation} f_r(h,t) = || h+r-t||_\frac{1}{2}\end{equation}
As TransE involves simple translation operation by forcing one to one mappings, it does not perform well with multirelational graphs.

\item DistMult: The algorithm uses bilinear diagonal modeling using the trilinear dot product scoring function~\cite{Distmult} defined as follow: \newline \begin{equation}f_r(h,t) = h^Tdiag(r)t = \sum\limits_{i=0}^p [r]_i\cdot[h]_i\cdot[t]_i \end{equation} \begin{equation}  \newline where \hspace{0.5cm}  h,r,t \in \mathbb{R}^d \hspace{0.5cm} and \hspace{0.5cm} p = d-1\end{equation} Here h, r and t are embeddings of head, relationship and tail respectively. The score function has its own limitation as it deals with the symmetric relationships as the score function is only able to capture the pairwise interaction between components of h and t along the same dimensions ~\cite{KGE}.

\item ComplEx: This model represents entities and relationships as complex vector embeddings, which consist of two vectors, the real and imaginary~\cite{complex}. Fact assertion using the asymmetric score function defined below for fact (h,r,t): \begin{equation} f_r(h,t) = Re(h^Tdiag(r)\overline{t}) = Re(\sum\limits_{i=0}^p [r]_i\cdot[h]_i\cdot[\overline{t}]_i)\end{equation} \begin{equation}where \hspace{0.5cm}  h,r,t \in \mathbb{C}^d \hspace{0.5cm} and \hspace{0.5cm} p = d-1\end{equation} Also $\overline{t}$ is conjugate of t and $Re()$ results in providing the real part of the complex value. The algorithm is an extension of DistMult that involves complex embeddings leading to better modelling of asymmetric relations due to asymmetric scoring function. Now embeddings of h,r, and t no longer exist in real space, but complex space $\mathbb{C}$. For symmetric relationships, DistMult performs well with symmetric relations and ComplEx works well with antisymmetric relations~\cite{distma}.

\item HolE: Stands for Holographic embeddings~\cite{Hole} which associate each entity with a vector to capture its latent semantics~\cite{KGE} The scoring function for HolE is defined as: \begin{equation}f_r(h,t) = r^T(h*t) = \sum\limits_{i=0}^p [r]_i \sum\limits_{k=0}^p [h]_k \cdot [t]_z \end{equation} Where the circular correlation operator is defined as: \begin{equation}[h*t]_i = \sum\limits_{k=0}^p [h]_k \cdot [t]_z \end{equation} \begin{equation} where \hspace{0.5cm}  h,r,t \in \mathbb{R}^d \hspace{0.5cm} and \hspace{0.5cm} p = d-1\end{equation} HolE uses the simplicity of TrasnE and expressive power of the tensor product, to generate better embeddings. It is able to deal with asymmetric relationship due to the fact that circular correlation is not commutative i.e. $h*t \neq t*h$~\cite{KGE}.

\item ConvE: Due its neural network structure it performs better in making non-linear transformations~\cite{ConvE}. It uses 2D convolution which is better at extracting interatcions between embeddings as compared to 1D convolution network. The scoring function is defined as: \begin{equation}f_r(h,t) = f(vec(g[\bar{e_s};\bar{e_r}]*w))W)e_0\end{equation} Here, g is the non-linear activation function, vec indicated the 2D reshaping function and * is the linear convolution operator and W is the weight matrix. Also, $\bar{e_s}$ and $\bar{e_r}$ are 2D reshaping of head entity embedding and the relationship embedding respectively, with loss function BCE used :
  \begin{equation}
    L(p,t) = -\frac{1}{N}\sum_i(t_i\cdot log(p_i) + (1-t_i)\cdot log(1-p_i) ) 
  \end{equation}
  
\item ConvKB: Unlike ConvE it utilizes 1D convolution and models the relationship among same dimensional entries of the embeddings~\cite{ConvKB}, which leads it to generalize the translational characteristics present in the translation based models. It represent a k-dimension embedding of every triple $(v_h,v_r,v_t)$ which is viewed as a matrix $A = [v_h,v_r,v_t]$. An operation is performed on each row of the generated matrix, the row i can be represented as $A_i,_: \in \mathbb{R}^{1 \times 3}$. An additional filter $\omega \in \mathbb{R}^{1 \times 3}$ is operated on every row to examine the global relationships, which enhances the translation characteristics of algorithm. 

Finally the feature map v is generated, $v = [v_1,v_2...v_k] \in \mathbb{R}^k$ which is generated as follows:
  \begin{equation}
    v_i = g(\omega\cdot A_i,_: + b)   \hspace{0.5cm}
    where \hspace{0.2cm} b \in \mathbb{R}
  \end{equation}
  Finally, these feature maps are used in scoring function of ConvKB where $*$ is convolution operation $\Omega$ and $w$ are shared parameters.
  \begin{equation}
    f_r(h,t) = concat(g([v_h,v_r,v_t]*\Omega))\cdot w
  \end{equation}
\end{itemize}

\subsection{Visualizing Knowledge Graph}
The below visual represents a section of original KEGG knowledge graph where different entities are denoted using different colours. It can observed how few entities are acting as the bridge to connected multiple clusters in the node.
It is clear through visualization that the number of connection per entity is highly imbalanced, with few nodes connected to
a large section of other nodes creating a cluster of their own.
\begin{figure*}[h!]
  \centering
  \includegraphics[width=0.7\linewidth]{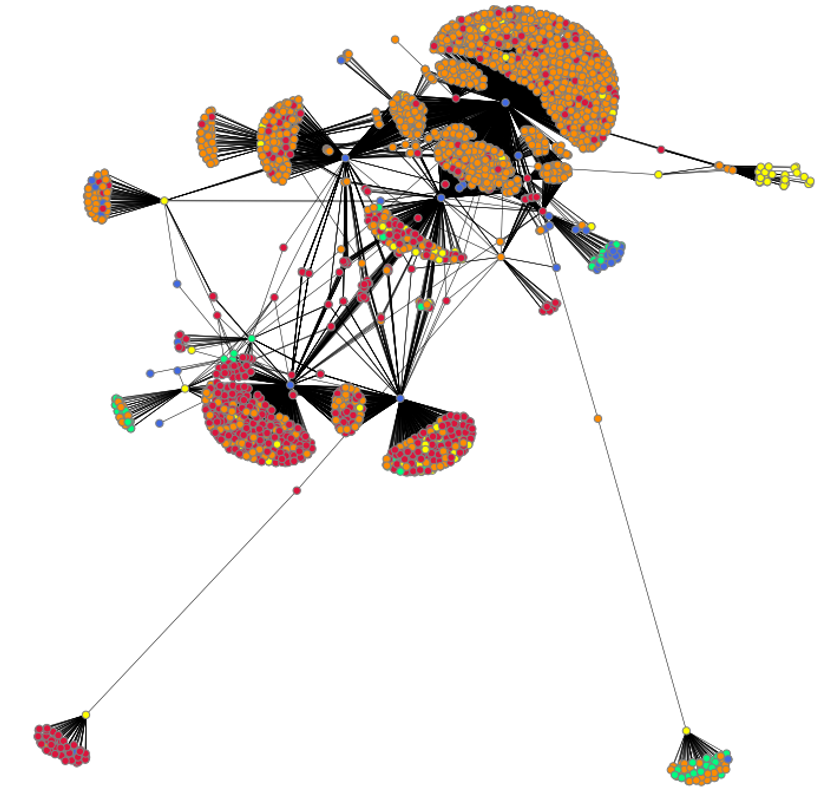}
  \caption{A section of KEGG Knowledge Graph}
  \label{fig:logo}
\end{figure*}

\subsection{Novel Node Similarity Measure Using Graph Embedding and Link Prediction Techniques}
We will use the cosine similarity measure between embeddings and information gathered from the link prediction task to improve the performance of the similarity measure by capturing various aspects of similarity.
The link prediction outputs are transformed into probabilities using calibration where
higher link prediction probability represents higher chances of a link existing between those two
nodes. We exploit the fact that if two drugs are similar their interaction
with other entities in the dataset should be similar i.e. their chance of linking with other
drugs should follow similar trends. Our aim will be to find some
kind of measure that can provide real value output to show the extent of similarity between two
entities, for our use case these two entities will be two drugs so our system will act as a drug
similarity system.
The cosine similarity between two graph embeddings, each of length m defined as $V_1 = [v_{11},v_{12}....v_{1m}]$ and $V_2 = [v_{21},v_{22}....v_{2m}]$ is defined as follows:
\begin{equation}\cos(V_1,V_2) = \frac{\sum\limits_{1}^m v_{1i} \times v_{2i}}{\sqrt{\sum\limits_{1}^m v_{1i}^2}\times\sqrt{\sum\limits_{1}^m v_{2i}^2}}\end{equation}
A loss function is defined to measure the difference between the link prediction
results of any two drugs with respect to all other possible entities in the dataset. \begin{equation} MSE = \sum\limits_{i=1}^{|E|}\frac{(p_{Ai} - p_{Zi})^2}{|E|} \end{equation} Here $p_{Ai}$ represents link prediction probability of drug A with entity i and $|E|$ stands of count of entities taken into account in this activity. 
Finally, MSE will give us a sense of the difference
in interaction between two drugs with other entities. The final similarity measure will use information from both similarity measures using graph embeddings and the link prediction loss function results. The novel similarity measure is defined in next equation.
\begin{equation}Sim(d_1,d_2) =  cos(V_1,V_2)/ MSE\end{equation}

Here, MSE is link prediction loss, $V_1$ and $V_2$ are embeddings from drugs $d_1$ and $d_2$ and Sim represents the overall similarity score between a pair of drugs.

\section{Evaluation}
\subsection{Model Settings}
Like any other machine learning model, the
performance of embeddings is dependent on the quantity and quality of data used, different hyperparameters. Our experiments use embedding sizes of 100, 200, 300 with loss functions Multiclass NLL Loss and Binary Cross Entropy Loss. Optimizers used are Adam (Adaptive Moment Estimation) and Stochastic
Gradient Descent are considered with learning rates 1e-1 and 1e-3 for experimentation. The test includes regularization techniques that are L1 (Lasso
Regression), L2 (Ridge Regression) and L3 (Nuclear 3-norm) proposed in the ComplEx-N3 paper~\cite{L3} with regularization constants 1e-5, 1e-2 and 1e-1.
There are other hyperparameters which
are valid for convolution graph neural net based models like ConvE and ConvKB. Their
parameters are kept fixed as number of feature maps per convolution kernel as 32, convolution
kernel size as 3, dropout at embedding layer, convolution layer and dense layer as
0.2, 0.3 and 0.2 respectively.

\subsection{Performance of Algorithms on Link Prediction Task}
The six different graph embedding model are trained on KEGG data which is split in 80:20 as train/test set and evaluated on the link
prediction task using measures like MRR and Hits@k. 
\begin{figure*}[h!]
  \centering
  \includegraphics[width=0.6\linewidth]{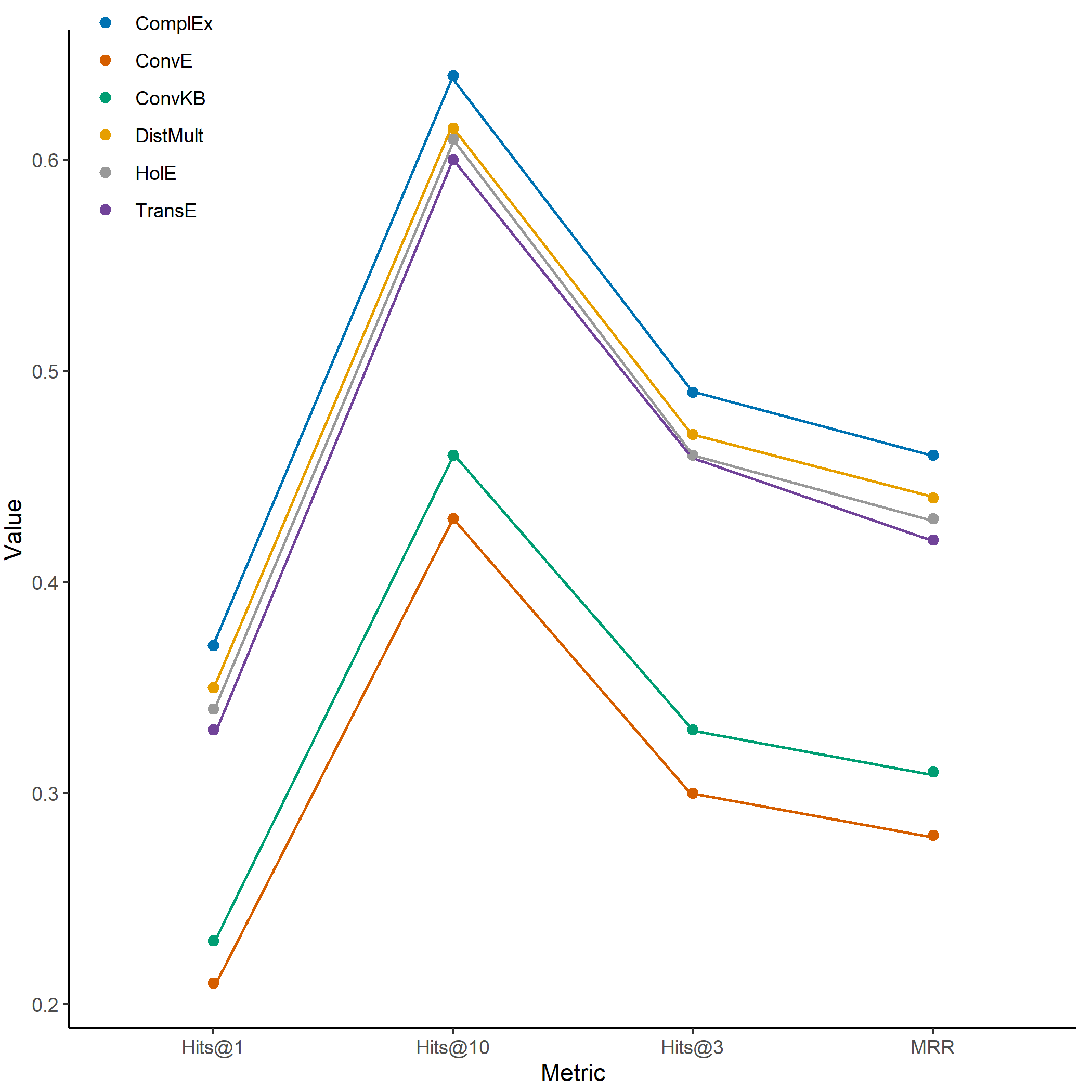}
  \caption{Performance of different graph embedding algorithm on KEGG}
  \label{fig:logo}
\end{figure*}
ComplEx has performed better than all other algorithms on the link
prediction task providing the best MRR and Hits@k scores. Surprisingly, the complex graph
convolution network models like ConvE and ConvKB have performed poorly on the KEGG
dataset.
Overall, ComplEx algorithm with embedding size of 300, Adam optimizer with learning rate of 1e-3 with Multiclass NLL Loss and L3 Regularization technique with 1e-2 as regularization constant provides best scores like 0.46 as MRR, 0.64, 0.49 and 0.37 as Hits@10, Hits@3 and Hits@1 respectively.

\subsection{Visualizing Different Entity Type in 2D Embedding Space}
The dimensionality reduction method PCA (Principal Component Analysis) \cite{PCA} is used
to reduce the larger space embeddings to smaller spaces that are easier to visualise. The embeddings are transformed to 2D space and
visualized on a plot marking the different entities present in dataset namely drug, disease, gene,
network and pathway with different colours for better understanding of relationships between
entities.
It can be observed how the graphed embedding algorithm cleverly generated the embeddings
such that different entities are present in the different sections of the plot, storing their personalities
and entity type’s within them.
\begin{figure*}[h!]
  \centering
  \includegraphics[width=0.5\linewidth]{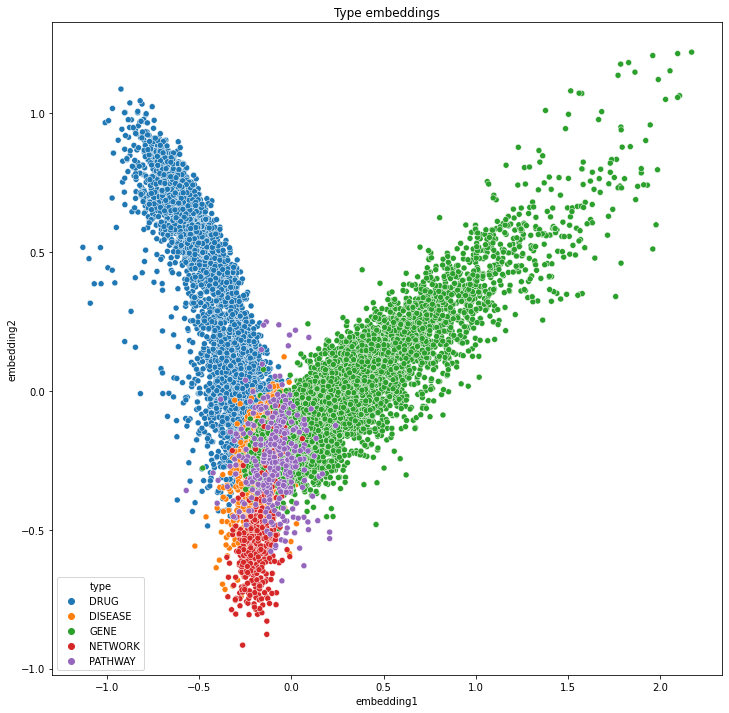}
  \caption{Plotting of embeddings in 2D spaces marked wrt to entity type}
  \label{fig:logo}
\end{figure*}

\subsection{Insights from Link Predictions}
The learned model is used to find possibilities for new unseen relationships between entities to provide
meaningful insights.
The model outputs the rank
of the statement, score of the statement which is normalized to a definite range from 0 to 1
to give a sense of a comparative probabilities among different statements.
The transformation of the scores
(real number) to probabilities (between 0 and 1) is performed using the expit transformation,
which takes any real number x and transforms it to a value in (0, 1). The model used for below predictions is the ComplEx model that has provided best scores, 0.46 as MRR, 0.64, 0.49 and 0.37 as Hits@10, Hits@3 and Hits@1 respectively.

\begin{table}
\caption{Calculating probability of relationships using the learned ComplEx Model}\label{tab1}
\begin{tabular}{|p{8cm}|p{1.3cm}|p{1.3cm}|p{1.3cm}|}
\hline
{\bfseries Statement} & {\bfseries  Rank} & {\bfseries Score} & {\bfseries Prob}\\
\hline
hsa04024 PATHWAY\_GENE HSA:51196 & 236 & 3.704783 & 0.975985 \\
\hline
D11034 DRUG\_EFFICACY\_DISEASE H00409 & 2 & 4.851221 & 0.992242 \\
\hline
D04905 DRUG\_TARGET\_PATHWAY hsa05010 & 1 & 4.979891 & 0.993172 \\
\hline
N00060 NETWORK\_GENE HSA:23401 & 1 & 5.399962 & 0.995504 \\
\hline
hsa04024 PATHWAY\_GENE HSA:6336 & 19814 & -0.21121 & 0.447393 \\
\hline
D11056 DRUG\_TARGET\_GENE HSA:7388 & 7636 & 0.089212 & 0.522288 \\
\hline
D11056 DRUG\_TARGET\_GENE HSA:3352 & 133 & 2.823472 & 0.943931 \\
\hline
N00399 NETWORK\_GENE HSA:9217 & 27812 & -0.44892 & 0.389616 \\
\hline
D04905 DRUG\_TARGET\_PATHWAY hsa04728 & 25 & 4.369776 & 0.987504 \\
\hline
N00399 NETWORK\_GENE N00399 & 16476 & -0.05818 & 0.485458 \\
\hline
H00242 DISEASE\_GENE D11034 & 28037 & -0.20788 &  0.448214 \\
\hline
N00060 DRUG\_TARGET\_PATHWAY hsa04380 & 32017 & -0.92776 & 0.283378 \\
\hline
\end{tabular}
\end{table}
The table shows that there are a few assertions for which the model projected fairly high odds.
The probability of relationship between D11056 (Drug name : Mirtazapine hydrate)
and HSA:3352 (Gene name : HTR1D, 5-HT1D, HT1DA, HTR1DA, HTRL, RDC4) under this model is
quite high with 0.94, which suggests a possible connection between these two entities.
 Similarly,
the relationship between D04905 (Drug name : Memantine hydrochloride) and pathway hsa04728 have a high connection probability of 0.987 under the model.

\subsection{Drug Similarity System Using Graph
Embeddings and Link Prediction}
Graph embeddings generated by the trained graph model are utilized to determine the possible similarity of two drugs. A loss
function like mean square error is used to measure the difference between the link prediction
scores of drug A with other entities and drug B with other entities. The new similarity measure incorporates inputs from the cosine
similarity between the embeddings of the pair of drugs and the link prediction score mean square error, which is expressed in section 4.5.
The top 10 possibly similar drugs with respect to D00043 (Drug name : Isoflurophate Fluostigmine) sorted based on the novel similarity measure are defined below:

\begin{table}
\caption{Top 10 possibly similar drugs with respect to Drug D00043 (Drug name : Isoflurophate Fluostigmine)}\label{tab1}
\begin{tabular}{|p{0.5cm}|p{1.1cm}|p{1.5cm}|p{1.5cm}|p{1.7cm}|p{5.4cm}|}
\hline
{\bfseries No.} & {\bfseries  Drug} & {\bfseries Cosine} & {\bfseries MSE} & {\bfseries Ratio} & {\bfseries Drug Name}\\
\hline
1 & D01228 & 0.933001 & 0.000289 & 3223.509127 & Distigmine bromide (JP18/INN)\\
\hline
2 & D00196 & 0.939603 & 0.000306 & 3075.463914 & Physostigmine (USP) \\
\hline
3 & D03751 & 0.943170 & 0.000324 & 2915.059617 & Icopezil maleate (USAN)\\
\hline
4 & D02418 & 0.944881 & 0.000328 & 2884.443301 & Physostigmine salicylate (JAN/USP)\\ 
\hline
5 & D06288 & 0.918739 & 0.000339 & 2706.522993 & Velnacrine maleate (USAN)\\
\hline
6 & D00469 & 0.928133 & 0.000347 & 2677.510252 & Pralidoxime chloride (USP)\\
\hline
7 & D05981 & 0.934112 & 0.000366 & 2555.506934 & Suronacrine maleate (USAN)\\
\hline
8 & D02558 & 0.957066 & 0.000386 & 2479.764194 & Rivastigmine tartrate\\
\hline
9 & D03826 & 0.934139 & 0.000388 & 2404.652320 & Physostigmine sulfate (USP)\\
\hline
10 & D02068 & 0.915335 & 0.000388 & 2357.032571 & Tacrine hydrochloride (USP)\\
\hline
\end{tabular}
\end{table}

It can be observed that D01228 (Drug name : Distigmine bromide (JP18/INN)) provides the highest ratio score with respect to drug
D00043 (Drug name : Isoflurophate Fluostigmine) due to its high cosine similarity and low
link prediction mean square error, indicating a higher probability to perform similar to D00043
when interacted with other possible entities.
Both drugs belong to the Neuropsychiatric agent class and are part of drug groups
DG01595 (Drug group name : Cholinesterase inhibitor) and DG01593 (Drug group name :
Acetylcholinesterase inhibitor) and targeting similar genes like HSA:43 (Gene name : ACHE,
ACEE, ARACHE, N-ACHE, YT), indicating a high similarities between the pair of
drugs.
For evaluation we have used Tanimoto coefficient, which is used to calculate the chemical similarity between molecules. It is defined as below, where S represent molecular similarity between A and B, a represents the number of on bits in A, b is number of on bits in B and c represents number of on bits in both A and B.
\begin{equation}
S_{A,B} = \frac{c}{a+b-c}
\end{equation}
The Tanimoto coefficient values for top drugs are 0.049, 0.056, 0.01, 0.046, 0.014, 0.021, 0.012, 0.77, 0.53. The chemical structure similarity is not capable to generalize the trend as there are drugs that treat same clinical problems but are different in structures, such as Migitol and Glipizide  which both are used for diabetes but have completely different structures~\cite{simi}. Finally some clinical drug similarity experiments will be a good choice for evaluation which is kept out of scope from this research.

\section{Conclusion}
Overall, the use of graph embedding models is an excellent choice to tackle the problem of finding similarity of drugs, and can be easily scalable by incorporating other medical datasets to provide our graph embedding model with more connected and larger databases. More data which eventually lead our models to generalize better by producing less over-fitted model, eventually providing better performance on link prediction task.
Graph embeddings models using neural networks, convolution neural networks and attention models is an exciting field which can help
us learn better about the complex graph structure their nodes and edges. Representing entities
and relationships of a knowledge graph in a form of low dimensional embedding can be used
for various graphical structures to find new insights using operations like link predictions. This work effectively contributes a machine learning pipeline that uses graph embeddings and link prediction algorithms to uncover drug similarity and capture unique relationships and possibilities in a biomedical database, with systematic comparison between different graph embedding algorithms. 

%
%
%
%

\end{document}